\documentclass[aps,prl,twocolumn,superscriptaddress,preprintnumbers,showpacs,floatfix]{revtex4}
\usepackage{graphics,colordvi}
\def\eq#1\en{\begin{equation} #1 \end{equation}}

\def\eqa#1\ena{\begin{eqnarray} #1 \end{eqnarray}}

\newcommand{\GeV}{\mbox{\rm GeV}}

\usepackage{graphicx,color}

\def\pp#1{\partial_{#1}}

\begin{document}

\title{Spacetime noncommutativity and ultra-high energy cosmic ray experiments}
\author{Raul Horvat}
\author{Dalibor Kekez}
\author{Josip Trampeti\'{c}}
\affiliation{Physics Division, Rudjer Bo\v skovi\' c Institute, Zagreb, Croatia}


\begin{abstract}
If new physics were   
capable to push the  neutrino-nucleon inelastic cross
section three orders of magnitude beyond the standard-model (SM) 
prediction, then
ultra-high energy (UHE) neutrinos would have already been observed at 
neutrino observatories. 
We use such a constraint to reveal information on the scale
of noncommutativity (NC) $\Lambda_{\rm NC}$ in 
noncommutative gauge field theories (NCGFT) where
neutrinos possess a tree-level coupling to photons in a
generation-independent manner. In the energy range of interest ($10^{10}$ to
$10^{11}$ GeV) the $\theta$-expansion ($|\theta| \sim 1/\Lambda_{\rm NC}^2$) 
and therefore the perturbative expansion in terms of $\Lambda_{\rm NC}$ retains 
no longer its meaningful character, 
forcing us to resort to those NC field-theoretical frameworks involving the 
full $\theta$-resummation. Our
numerical analysis of the contribution to the process coming from the photon
exchange, pins impeccably down a lower bound on $\Lambda_{\rm NC}$ 
to be as high as around up to 900 (450) TeV,
depending on the estimates for the cosmogenic neutrino flux. 
If, on the other hand, one considers a surprising recent 
result occurred in Pierre
Auger Observatory (PAO) data, that UHE cosmic rays are mainly composed of
highly-ionized Fe nuclei, then our bounds get weaker, due to the diminished
cosmic neutrino flux. Nevertheless, we show that even for the very high
fraction of heavy nuclei in primary UHE cosmic rays, our method may still
yield remarkable bounds on $\Lambda_{\rm NC}$, typically always above 
200 TeV. Albeit, in this case  
one encounters a maximal value for the Fe
fraction from which any useful information on $\Lambda_{\rm NC}$ can be drawn,
delimiting  thus the applicability of our method.
\end{abstract}

\maketitle



The observation of ultra-high energy (UHE) neutrinos from extraterrestrial
sources would open a new window to look to the cosmos, as such neutrinos may
easily escape very dense material backgrounds around local astrophysical
objects, giving thereby information on regions that are otherwise hidden to
any other means of exploration. In addition, neutrinos are not deflected on
their way to the earth by various magnetic fields, pointing thus back to
the direction of distant UHE cosmic-ray source candidates. This could also
help resolving the underlying acceleration in astrophysical sources.
    
Since the GZK-structure in the energy spectrum of UHE 
cosmic rays at $\sim 4 \times 10^{19}$ eV 
has been observed recently with high statistical accuracy \cite{Aglietta:2007yx}, 
the flux of the so-called 
cosmogenic neutrinos, arising from photo-pion production on the
cosmic microwave background $p \gamma_{CMB} \rightarrow \Delta^{*}
\rightarrow N\pi$ 
and subsequent pion decay, is now guaranteed to exist. 
Although estimates for the cosmogenic neutrino flux are very
model-dependent, primarily due to our insufficient knowledge of the nature
and the origin of UHE cosmic rays, possible ranges for the size of the
flux of cosmogenic neutrinos can be obtained from separate analysis of the data
from various large-scale observatories \cite{Fodor:2003ph,Protheroe:1995ft}. 
With this knowledge of
cosmogenic neutrinos and the increased exposure of large-scale detectors,
a tight model-independent constraint on the neutrino-nucleon cross section 
can be placed \cite{Anchordoqui:2004ma}. 
A similar analysis but with a different approach to the problem 
was carried out in \cite{Barger:2006ky}. 

We note that the uncertainty in the flux of cosmogenic neutrinos may be further
aggravated in view of apparently contradictory recent results regarding the
chemical composition of UHE cosmic rays. While the
Pierre Auger Observatory (PAO) data indicate that
UHE cosmic rays may be highly-ionized Fe nuclei \cite{Abraham:2010yv},
the results from the High Resolution Fly's Eye Collaboration
(HiRes) are consistent with the proton dominance in the UHE
region \cite{Abbasi:2009nf}. In the former case the flux of cosmogenic
neutrinos would possibly be much smaller than that in the latter case
\cite{Lemoine:2004uw}. Further confusion was
brought in by the fact that both experiments are intrinsically inconsistent
regarding the arrival directions of UHE cosmic rays \cite{Abbasi:2010xt}.
Note also that the most recent evidence of anisotropy of UHE cosmic-ray arrival 
directions at the 99\% confidence level \cite{:2010pk} by 
the PAO would question further its interpretation that primary 
UHE cosmic rays are likely to be dominated by heavy nuclei.
On top of that, there is another inconsistency besetting the PAO data: the
average of the depth in the atmosphere at which the
shower contains the largest number of high energy particles implies an Fe
fraction $< 60 \%$, whereas the analysis of its standard deviation implies for
the same fraction to be $> 90\%$.
Although previously most observations and arguments seemed to favor protons
as a dominant component at the highest energies, and although
some speculations exist that inconsistencies in the
PAO data rather point to some new physics than to a heavy
component in the UHE cosmic rays \cite{Wilk:2010iz},
this issue still remains an open question \cite{Schwarzschild:2010zz}.

In this paper, we employ the aforementioned upper bound \cite{Anchordoqui:2004ma} 
on the cross section to constrain the scale of noncommutativity entering
NCGFT in which neutrinos couple directly to photons. We
use upper bounds derived by non-observation of UHE
neutrino-induced events in RICE Collaboration search results \cite{Kravchenko:2002mm} and
exploiting the predictions for the cosmogenic neutrino flux 
from \cite{Fodor:2003ph,Protheroe:1995ft}.  
Also, to realize the second possibility as indicated by the PAO data, we
derive our bounds by using the same predictions for the cosmogenic neutrino
flux, but now diminished by the fraction of FE nuclei in the UHE portion of
cosmic ray spectrum, which we treat here as a free parameter.

NCGFT that we consider is based on Moyal-Weyl star($\star$)-product 
\footnote{The Moyal-Weyl star($\star$)-product of some fields $f,\;g$,
$(f \star g) (x) = f(x) e^{\frac{i}{2}\stackrel{\leftarrow}{\partial_\mu}
\theta^{\mu\nu}\stackrel {\rightarrow}{\partial_\nu}} g(x)$, 
is a non-local bilinear expression in the fields and their derivatives, and it is associative 
but, in general, not commutative. 
It takes the form of a series 
in a Poisson antisymmetric tensor $\theta^{\mu\nu}$ defined via $\star$-commutator, 
$[x^\mu,x^\nu]_{\star}=x^\mu \star x^\nu - x^\nu \star x^\mu =i\theta^{\mu\nu}$,
which implements the spacetime NC.
Also, the important property is 
$\int d^4x\; f\star g \star h 
=\int d^4 x\; f \cdot (g \star h)$, for $\theta$ = constant.}
and it belongs to a 
class of models that expand the NC action in $\theta$ \emph{before}
quantization~\cite{Madore:2000en,Schupp:2002up}. 
In these models Seiberg-Witten (SW) maps \cite{Seiberg:1999vs} are necessary to express
the noncommutative fields $\widehat \psi$ and $\widehat A_\mu$, that appear in the action
and transform under NC gauge transformations, in terms
of their asymptotic commutative counterparts $\psi$ and $A_\mu$. The algebra
generated here is the enveloping algebra. 
Computation of any (physical) quantities, obtained in this approach, 
are given as a power series in the NC parameter $\theta$, 
thus showing no UV/IR mixing on the quantum level.

Admittedly, in other approaches, with $\theta$-unexpanded, 
i.e., models based on $\star$-products only \cite{Chaichian:2001py}, 
the celebrated quantum-gravity 
effect shows up: the UV/IR mixing 
\cite{Martin:1999aq,Abel:2006wj,Schupp:2008fs}. 
The same would happen in our approach. 

Coupling of matter fields to Abelian gauge bosons is a NC
analogue of the usual minimal coupling scheme. 
It will imply only a vector-like NC coupling between photons and neutrinos in 
the U(1)$_{\star}$ gauge-invariant way, 
thus endowing also otherwise sterile right-handed 
(RH) neutrino components a new interaction.

It is possible to extend the model \cite{Schupp:2002up} to the NC electroweak
model based on the other gauge groups, like NC version of the 
standard model (NCSM), 
etc. \cite{Calmet:2001na,Aschieri:2002mc}. 
That NCSM \cite{Calmet:2001na} at $\theta$-order does break Lorentz symmetry, 
appears to be anomaly free \cite{Martin:2002nr}
and it does have remarkable well behaved one-loop quantum corrections \cite{Buric:2006wm},
as well as some other models do \cite{Bichl:2001cq,Latas:2007eu,Ettefaghi:2007zz}.

Signatures of noncommutativity and/or the bounds on the NC scale come from
neutrino astrophysics \cite{Schupp:2002up,Minkowski:2003jg,Haghighat:2009pv} 
and cosmology \cite{Akofor:2008gv,Horvat:2009cm}; from high energy particle
physics \cite{Melic:2005hb,Buric:2007qx,Alboteanu:2007bp}, producing a
scale of noncommutativity of order few TeV's.
Typical low energy nonaccelerator experiments are the Lamb-shift \cite{Chaichian:2000si}, 
and clock-comparison
experiments \cite{Carroll:2001ws}.
The bound from $Z \rightarrow \gamma\gamma$ decay on $\Lambda_{\rm NC}$,
of order a few TeV's \cite{Buric:2007qx}, is the most robust due to the finite one-loop
quantum corrections in the gauge sector of the nmNCSM \cite{Buric:2006wm}.
Also, the interesting upper bound on the scale of NC, of order of one
tenth of the Planck mass, can be obtained from the
black-hole physics in the NC spacetime \cite{Nicolini:2005vd}.

Neutrinos do not carry an electromagnetic charge
and hence do not couple directly to photons. 
However, in the presence of spacetime NC, it is possible to couple neutral particles to 
photons via a $\star$-covariant derivative $\widehat D_\mu \widehat\psi$.
Thus, the action for neutrinos that couples to an Abelian gauge boson 
in the NC background, invariant under the NC gauge transformations, is
\begin{eqnarray}
S_{\rm NC} &=& \int d^4 x \left(\,\overline{\widehat\psi} \star 
i\gamma^\mu\widehat D_\mu \widehat\psi
-m \overline{\widehat\psi} \star \widehat\psi\right)\,,
\label{NCa}\\
\widehat D_\mu \widehat\psi&=&\partial_\mu \widehat\psi - 
i\kappa e [\widehat A_{\mu}\stackrel{\star}{,} \widehat\psi],
\label{ncd}\\
\widehat A_\mu &=& A_\mu + e\theta^{\nu\rho}A_{\rho}
[\partial_{\nu}A_{\mu}-\frac{1}{2}\partial_{\mu}A_{\nu}] + {\cal O}(\theta^2)\,, 
\label{A} \\
\widehat \psi &= &\psi + e\theta^{\nu\rho}A_{\rho}\partial_{\nu}\psi + {\cal O}(\theta^2)\,,
\label{psi}
\end{eqnarray}
where (\ref{A}) and (\ref{psi}) are, up to the first order in $\theta$, ordinary SW maps 
for the Abelian NC gauge potential 
and ``chiral'' SW map for the NC ``chiral'' neutrino field, respectively
\footnote{Note that a ``chiral'' Seiberg-Witten map is compatible with grand unified models 
where fermion multiplets are chiral \cite{Aschieri:2002mc}.
In eqs (\ref{NCa},\ref{ncd}) with chiral fermion fields, one may think of the NC
neutrino field $\widehat \psi$ as having left charge $+\kappa e$, right charge $-\kappa e$
and total charge zero.
Coupling $\kappa e$ corresponds to a multiple (or fraction) $\kappa$ of the
electric charge $e$.
From the perspective of non-Abelian gauge theory,
one could also say that the neutrino field is charged in NC
analogue of the adjoint representation with the matrix multiplication replaced by the $\star$-product. 
For this model where only the neutrino has dual L/R charges,
$\kappa = 1$ is required by the gauge invariance of the action.}.
Here, we fix the NC scale by defining $\theta^{\mu\nu} \equiv
c^{\mu\nu}/\Lambda_{\rm NC}^2 $, such that the matrix elements of $c$ are of
order one.

Performing the SW and $\star$-product expansions up to the 
first order in $\theta$
one obtains the following electromagnetically gauge-invariant NC action 
for photons and neutrinos in terms of commutative fields and parameters \cite{Schupp:2002up}:
\begin{eqnarray}
S_{\rm NC} &=&-\frac{e}{2}\int d^4 x  
\frac{}{}
\bar \psi \,A_{\mu\nu}\left(
i \theta^{\mu\nu\rho}\pp\rho -\theta^{\mu\nu}m\right)\psi\,,
\label{SWaction}
\end{eqnarray}
with ${\theta}^{\mu\nu\rho}=
{\theta}^{\mu\nu}\gamma^{\rho}+{\theta}^{\nu\rho}\gamma^{\mu}+
{\theta}^{\rho\mu}\gamma^{\nu}$; $A_{\mu\nu}=\partial_\mu A_\nu-\partial_\nu A_\mu$.
From (\ref{SWaction}) we extract the Feynman rule for massless left/right(L/R)-handed neutrinos
(the same for each generation and $k = k' + q$) \cite{Schupp:2002up}: 
\begin{eqnarray}
{\Gamma}^{\mu}_{\rm L \choose \rm R}(\bar\nu \nu \gamma )
&=&\frac{i}{2}e(1 \pm \gamma_5){\theta}^{\mu\nu\rho}k_{\nu}q_{\rho}\;.
\label{expFR}
\end{eqnarray}

The double differential deep inelastic scattering cross 
section $d^2 \sigma_{\rm NC}
/dx dy$ for the process $\nu N \rightarrow \nu \,+\,anything$ via the ${\bar\nu} \nu
\gamma$ vertex as given by (\ref{expFR}),  
can be calculated with the help of the structure functions,
\begin{eqnarray}
\frac{d^2\sigma_{\rm NC}}{dxdy}
&=&
{\cal I}\;
\frac{d^2\sigma_{\mbox{\rm\scriptsize comm}}}{dxdy}\,,
\label{sigma}\\
{\cal I}
&=&
\frac{1}{2\pi}
\int_0^{2\pi}\,d\varphi\, (\frac{kck^\prime}{2\Lambda_{\rm NC}^2})^2
\nonumber \\ 
&{}& \hspace{-2cm}
\approx\left((c_{01}-c_{13})^2+(c_{02}-c_{23})^2\right)\frac{E_\nu^3 M_N}{4\Lambda_{\rm NC}^4}
\,x\,y\,(1-y),\;\;
\label{intlin}\\
\frac{d^2\sigma_{\mbox{\rm\scriptsize comm}}}{dx\,dy}
&=&
\frac{2\pi\alpha^2}{E_\nu M_N (xy)^2}
\nonumber \\ 
&{}& \hspace{-2cm}
\times\left[ (1-y)F_2^\gamma + y^2 x F_1^\gamma + y(1-y/2) x F_3^\gamma \right]~.
\label{cscomm}
\end{eqnarray}
\noindent 
Here the Bjorken variable $x$ is given in terms of the inelasticity $y$, the
total energy of the system $s$ and the invariant momentum transfer $-Q^2$ as
$x = Q^2/(ys)$. The azimuth angle $\varphi$ in (\ref{intlin}) is defined with respect to the
direction of the incident neutrino. The quantity
$\sigma_{\rm comm}$ refers to the cross section a neutrino would have as if it
were a Weyl particle with a charge $e$.
The structure function $F_{i}$'s for the isoscalar nucleon are given by 
\begin{eqnarray}
2 x F_1^\gamma(x,Q^2) &=& F_2^\gamma(x,Q^2)~,
\label{DIS:F1gamma}
\\
\frac{1}{x} F_2^\gamma(x,Q^2)
&=&
\frac{1}{2}(Q_u^2+Q_d^2)
    [u(x,Q^2)+\bar{u}(x,Q^2)\nonumber\\&+&d(x,Q^2)+\bar{d}(x,Q^2)]
\label{DIS:F2gamma}
\nonumber \\
&+&
2 Q_c^2 c(x,Q^2) + 2 Q_s^2 s(x,Q^2)
\nonumber \\
&+&
2 Q_b^2 b(x,Q^2) + 2 Q_t^2 t(x,Q^2)~,
\\
F_3^\gamma(x,Q^2) &=& 0~.
\label{DIS:F3gamma}
\end{eqnarray}
and in our analysis 
we use the parton distribution functions  from the audibly available
package CTEQ4 \cite{Lai:1996mg}. The kinematical region of interest here is
that of high $Q^2$ and very small $x$ values
$x \approx 1.7 \times 10^{-7}/(E_{\nu}/10^{11}~\GeV)$,
because of the rapid increase of the parton densities towards the small
$x$'s. Different extrapolation approaches can result in uncertainties as
large as a factor of two at $E_{\nu} = 10^{12}~\GeV$ \cite{Kutak:2003bd}.     

Employing the upper bound on the $\nu N$ cross section derived from the RICE
Collaboration search results \cite{Kravchenko:2002mm} at $E_{\nu} =
10^{11}$ GeV ($4\times 10^{-3}$ mb for the FKRT neutrino flux \cite{Fodor:2003ph})), 
one can infer from (\ref{expFR}-\ref{DIS:F3gamma}) [also
employing $(c_{01}-c_{13})^2+(c_{02}-c_{23})^2=1$]    
on the scale of noncommutativity $\Lambda_{\rm NC}$
to be greater than 455 TeV, a really strong bound. One should however be
careful and suspect this result as it has been obtained from the conjecture
that the $\theta$-expansion stays well-defined in the kinematical region of
interest, and the more reliable limits on $\Lambda_{\rm NC}$ are expected to be
placed  precisely by examining low-energy processes \cite{Haghighat:2009pv}.
Although a
heuristic criterion for the validity of the perturbative $\theta$-expansion,
$\sqrt{s}/\Lambda_{\rm NC} \lesssim\;1$, with $s = 2 E_{\nu}M_N$, would
underpin our result on $\Lambda_{\rm NC}$, a more thorough inspection on
the kinematics of the process does reveal a  more stronger energy
dependence  $E_{\nu}^{1/2} s^{1/4}/ \Lambda_{\rm NC} \lesssim 1$.
In spite of an
additional phase-space suppression for small $x$'s in  
the $\theta^2$-contribution \cite{Alboteanu:2007bp} of
the cross section relative to the $\theta$-contribution, we find 
an unacceptably 
large ratio $\sigma({\theta^2})/\sigma({\theta}) \simeq 10^4$,
at $\Lambda_{\rm NC}=455$ TeV.
Hence,
the bound on $\Lambda_{\rm NC}$ obtained this way is incorrect, and our last resort 
is to modify the model adequately to include somehow the full-$\theta$
resummation, thereby allowing us to compute nonperturbatively in $\theta$.

The simplest possible modeling, 
which we propose here is: (a) Replacement of the NC fields with
the commutative ones,
$\widehat\psi\to\psi$ and $\widehat A_{\mu}\to A_{\mu}$; (b) Expansion of 
the $\star$-product up to all orders,
and resummation to the $\theta$-exact final results.
Following (a) and (b), the NC action 
instead of (\ref{NCa}) can be written now as
\begin{eqnarray} 
S_{\rm NC}(\theta) = -i e \int d^4 x \;\bar\psi\frac{}{}
\gamma^\mu (A_{\mu}\star \psi - \psi \star A_{\mu})\,.
\label{neaction}
\end{eqnarray}
At this point the above action corresponds to the 
neutrino-photon interaction proposed in Ref.
\cite{Chaichian:2001py}.
Next we expand the $\star$-product to all orders in $\theta$, 
and take derivatives on 
$\psi(x) \sim e^{ikx}\tilde\psi(k)$ and $A_{\mu}(x) \sim e^{iqx}\tilde A_{\mu}(q)$.
After getting the momentum dependence order by order we resume the obtained 
series into the nontrivial 
exponential phase factors $\exp{(\pm i\frac{q{\theta}k}{2})}$,
and extract a generation-independent Feynman rule for massless L/R neutrinos
\begin{eqnarray}
{\Gamma}^{\mu}_{\rm L \choose \rm R}(\bar\nu \nu \gamma )
&=&ie (1 \pm \gamma_5) \gamma^{\mu} \sin(\frac{q{\theta}k}{2})\;.
\label{sinFR}
\end{eqnarray}

By comparison of our Feynman rule (\ref{sinFR}) with those from 
\cite{Martin:1999aq,Schupp:2008fs},
it is clear that our model would also produce the  
UV/IR mixing in computing quantum corrections,
thus, our model is not perturbatively renormalizable and it is not clear how to
interprete the quantum corrections and to relate them to the observations \cite{Horvat:2010km}. 
In return, we obtain  
the $\theta$ well-behaved deep inelastic cross section 
at ultra-high energies.

With the aid of the full vertex (\ref{sinFR}), the neutrino cross section 
still retains the form (\ref{sigma}), but with the relevant quantity now defined as
\begin{eqnarray}
{\cal I}
&=&
\frac{1}{2\pi}
\int_0^{2\pi}\,d\varphi\, 4 \sin^2(\frac{kck^\prime}{2\Lambda_{\rm NC}^2})
\nonumber \\ &=&
2 \left( 1 - \cos(A) J_0(B)\right)~,
\label{avg}
\end{eqnarray}
\noindent where $J_0$ is the Bessel function of the first kind of
order zero, and
\begin{eqnarray}
A &=& \frac{E_\nu E_\nu^\prime}{\Lambda_{\rm NC}^2} c_{03}(\cos\vartheta-1)~,
\label{Acos}\\
B &=& \frac{E_\nu E_\nu^\prime}{\Lambda_{\rm NC}^2} \sin\vartheta
\nonumber \\
&\times&
\hspace{-2mm}
\mbox{\rm sign}(c_{01}-c_{03}) \sqrt{(c_{01}-c_{13})^2+(c_{02}-c_{23})^2}.
\label{Bsin}
\end{eqnarray}
The behavior of the cross section
with the scale of noncommutativity at fixed $E_{\nu} = 10^{10}$ GeV  and
$E_{\nu} = 10^{11}$ GeV, 
together with the upper bounds depending on the actual size of the
cosmogenic neutrino flux (FKRT \cite{Fodor:2003ph} and PJ \cite{Protheroe:1995ft}) 
as well as the total SM cross sections at these energies, are
depicted in our Figure 1. In order to maximize the effect of
noncommutativity, in our numerical calculations we choose 
$c_{01}-c_{13}=c_{02}-c_{23}=c_{03}=1$.

A few comments are in order regarding qualitative behavior of
the cross section. The common attribute for both energies is the
existence of a plateau at small $\Lambda_{\rm NC}$'s, where the cross section
tends to a constant value. Also, the oscillatory term in (\ref{avg})
enters the regime of
rapid oscillations for lower $\Lambda_{\rm NC}$'s for
both energies. This behavior can be read off from (\ref{avg}) by noting that the
argument of the cosine function grows as $\Lambda_{\rm NC}$ is decreasing
(for fixed energies), whilst the amplitude of oscillations as given by $J_0$
tends to zero asymptotically in the same limit.
The cross section for the upper energy does clearly surpass the measured
value below around 900 (450) TeV for the FKRT (PJ) flux, bringing us a
valuable information on the NC scale.
As far as the lower energy is concerned, the
upper bounds assigned to both fluxes are crossed, at energies around 350
(500) TeV, respectively. Thus, the conservative bound for the 
energy $E_{\nu} = 10^{10}$ GeV is around 350 TeV.  

If future data confirm that UHE cosmic rays are composed mainly of Fe
nuclei, as indicated, for the time being, by the PAO data, then still
valuable information on $\Lambda_{\rm NC}$ can be obtained with our method, as
seen in Fig. 2. Here we see the intersections of our curves with the
RICE results (cf. Fig.1)
as a function of the fraction $\alpha$ of Fe nuclei in the UHE cosmic rays.
We see that in all but one cases our method is capable to extract
information on $\Lambda_{\rm NC}$ even for Fe fractions as large as $\stackrel{>}{\sim} \rm 90 \%$,
giving always a conservative bound $\stackrel{>}{\sim} \rm 200$ TeV. The terminal point on
each curve gives the highest fraction of Fe beyond which our method becomes
unserviceable to draw any information on the NC scale. Thus, when more data
on Fe fraction become available, one can easily pinpoint the lower bound 
on $\Lambda_{\rm NC}$ with the aid of Fig. 2.

To summarize, we have used the upper bound on the neutrino-nucleon 
inelastic cross section, derived by non-observation of UHE neutrino-induced
events in extensive airshower arrays, to deduce information on the scale of
noncommutativity in NC gauge field theories. The fact that we
deal here with a process whose center-of-mass energy is three orders of
magnitude higher than that being achieved with terrestrial accelerators, 
has also implications for the theory itself. Namely, 
the dimensionless parameter involving the scale of noncommutativity and
entering the perturbative expansion shows a strong incident-energy 
dependence, such
that perturbative expansion at such UHE energies is no longer
meaningful. After having treated our model nonperturbatively in $\theta$, we
used such a model to derive a very strong bound on  the scale of
noncommutativity, at around 900 (450) TeV, depending on the model for the
total cosmogenic neutrino flux. Since the prediction for the NC
neutrino-photon vertex 
in a perturbative setting is quite robust amongst different models, we 
believe  that our $\theta$-resummed interaction is also a generic one. 
For those inclined to accept the PAO interpretation, we have shown that even
for the large fraction of Fe nuclei in the UHE cosmic rays 
($\stackrel{>}{\sim} \rm 90 \%$) one
may obtain remarkable bounds on the scale of noncommutativity $(\Lambda_{\rm NC}
\stackrel{>}{\sim} \rm 200 TeV)$.
Finally, one should keep in mind that
the limits derived here are conservative ones, meaning that any influence
of Fe nuclei as well as NC physics on the cosmogenic neutrino flux
(and consequently on the experimental upper limit on
the cross section) has not been treated.

\begin{figure}[top]
\begin{center}
\includegraphics[width=8.5cm,angle=0]{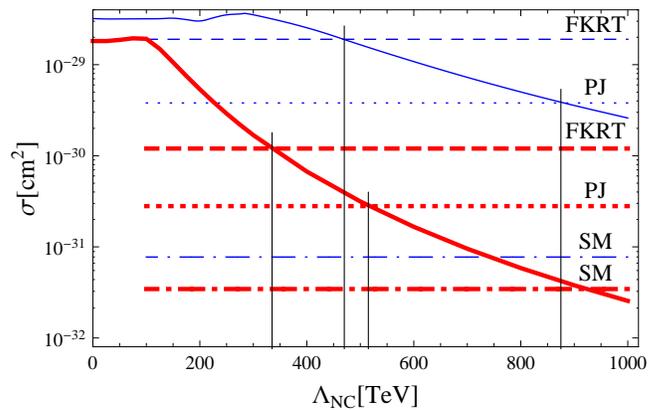}
\end{center}
\caption{$\nu N\to\nu\,+\, anything$ cross sections
vs. $\Lambda_{\rm NC}$ for $E_\nu=10^{10}~\GeV$ (red lines) 
and $E_\nu=10^{11}~\GeV$ (blue lines). FKRT and PJ lines are the upper
bounds on the neutrino--nucleon inelastic cross section, denoting different 
estimates
for the cosmogenic neutrino flux (see the text). SM denotes the SM total
(charged current plus neutral current) neutrino-nucleon inelastic cross
section. The vertical lines denote the intersections of our curves with the
RICE results.  }
\label{fig:ncSM-CrossSections}
\end{figure}

\begin{figure}[top]
\begin{center}
\includegraphics[width=8.5cm,angle=0]{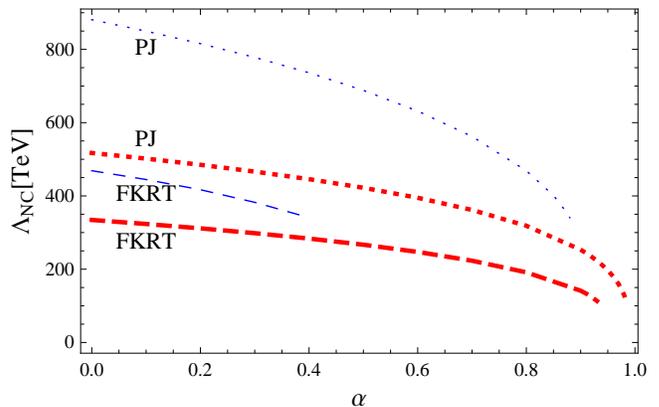}
\end{center}
\caption{The intersections of our curves with the RICE results (cf. Fig.1)
as a function of the fraction of Fe nuclei in the UHE cosmic rays. The
terminal point on each curve represents the highest fraction of Fe nuclei
above which no useful information on $\Lambda_{\rm NC}$ can be inferred with
our method.}
\label{fig:ncSM-LambdaVsAlpha}
\end{figure}

\section*{Acknowledgment}
The work of R.H., D.K and J.T. are supported by 
the Croatian Ministry of Science, Education and Sports 
under Contract Nos. 0098-0982930-2872 and 0098-0982930-2900, respectively. 
We would like to thank P. Schupp and J. You for useful discussions,
and W. Hollik at MPI for the hospitality.
The work of J.T. is in part supported by the HEPTOOLS 
under contract MRTN-CT-2006-035505.

\end{document}